\documentclass[conference]{IEEEtran}
\IEEEoverridecommandlockouts
\usepackage{cite}
\usepackage{amsmath,amssymb,amsfonts}
\usepackage{algorithmic}
\usepackage{graphicx}
\usepackage{textcomp}
\usepackage{xcolor}
\def\BibTeX{{\rm B\kern-.05em{\sc i\kern-.025em b}\kern-.08em
    T\kern-.1667em\lower.7ex\hbox{E}\kern-.125emX}}
\begin{document}

\title{Recursive Matrix Algorithms in Commutative Domain for Cluster with Distributed Memory 
\thanks{This work was completed with the support of RFBR grant No 16-07-00420.}}
\author{\IEEEauthorblockN{1\textsuperscript{st} Gennadi Malaschonok}
\IEEEauthorblockA{\textit{
 National University of Kyiv-Mohyla Academy, Kyiv, Ukraine }\\
Tambov State University,  Tambov, Russia  \\
malaschonok@gmail.com}
\and
\IEEEauthorblockN{2\textsuperscript{nd} Evgeni Ilchenko}
\IEEEauthorblockA{ 
\textit{Tambov State University}\\
Tambov, Russia \\
ilchenkoea@gmail.com}
}
\maketitle
\begin{abstract}  
We give an overview of the theoretical results for matrix block-recursive algorithms in commutative domains
 and present the results of experiments that we conducted with new parallel programs based on these algorithms on a supercomputer
 MVS-10P at the Joint Supercomputer Center of the Russian Academy of Science.
To demonstrate a scalability of these programs
we measure the running time of the program for a different number of processors and  plot the graphs of efficiency factor.
 Also we  present the main application areas in which such parallel algorithms are used.
It is concluded that this class of algorithms allows to obtain efficient parallel programs on clusters with distributed memory.
\end{abstract}
\section{Introduction}
J. Dongarra at his talk at International Congress ICMS-2016  \cite{Dongarra-2016} put attansion on the several difficult challenges.
He noted that the task of managing calculations on a cluster with distributed memory for algorithms with sparse matrices is today 
one of these the most difficult challenges. 

We have to add also one more problem.  
It is a high computational complexity, that can be connected with the type of the basic algebra: you can take a matrix over field or over commutative ring. 

For sparse matrices, it is not true that all computations over polynomials or integers can be effectivelly reduced due to the technic  of modular computations.
It was proved in theoretical investigatios of computational complexity for some algorithms with sparse matrices in  \cite{MVL}.
Below we give experiments with large sparse matrices, which confirm these theoretical reults.

We consider the class of block-recursive matrix algorithms. 
The most famous of them are standard and Strassen's block matrix multiplication, Strassen's block-matrix inversion  \cite{Strassen-1969}. 
 
 Block-recursive algorithms were not so important as long as the calculations were performed on computers with shared memory.  
 Only in the nineties it became clear that block-recursive matrix algorithms are required to operate with sparse large matrices
 on a supercomputer with distributed memory.
 
 Note that the generalization of Strassen’s matrix inversion algorithm \cite{Strassen-1969} with 
 additional permutations of rows and columns
by J. Bunch and J. Hopkroft \cite{Bunch-Hopkroft-1974} is not a block-recursive algorithm.
 
The block recursive algorithm for the solution of systems of linear equations
and for adjoint matrix computation which is some generalisation of Strassen's inversion in commutative domains was 
suggested in the papers \cite{1997}, \cite{2000} and \cite{2006}. See also at the book \cite{2002}.
However, in all these algorithms, except matrix multiplication, a very strong restriction are imposed on the matrix:
the leading corner minors should not be zero. 

This restriction was removed later.
The algorithm that computes the adjoint matrix, the echelon form, and the kernel of the matrix operator for 
the commutative domains was proposed in \cite{2008}.
The block-recursive algorithm for the Bruhat decomposition and the LEU decomposition for the matrix over the field was 
obtained in  \cite{2010}, and these algorithms were generalized for the matrices over commutative domains 
 in \cite{2013} and in \cite{2015}.

 In this article we review the main achievements in this class of algorithms and present the results of experiments
 that we conducted with these algorithms on a supercomputer
 MVS-10P at the Joint Supercomputer Center of the Russian Academy of Science.
 
In the next section, we present the main application areas in which such algorithms are used.

\section{Some important areas for applications of sparse matrices algorithms}
\subsection{ Computations of functions of  electronic circuits}
The behavior of electronic circuits can be described by Kirchhoff's laws.
The three basic approaches in this theory are direct current, constant frequency current and a current that varies with time. 
All these cases require the compilation and solution of sparse systems of equations (numerical, polynomial or differential). 
The solution of such differential equations by the Laplace method also leads to the solution of polynomial systems of equations
\cite{Paul-2001}.
\subsection{ Control systems}
In 1967  Howard H. Rosenbrock introduced  a useful state-space representation and transfer function matrix form for control systems, 
which is known as the Rosenbrock System Matrix \cite{Rosenbrock1967}.  Since that time, the properties of the matrix of polynomials being intensively studied in the literature of linear control systems.
%
\subsection{Computation of Gr\"obner bases}
Another important application is the calculation of Gr\"obner bases. A matrix composed of Buchberger S-polynomials is a strongly sparse matrix. 
Reduction of the polynomial system is performed when calculating the echelon and diagonal forms of this matrix.
The algorithm F4 \cite{1999} was the first such matrix algorithm.
%
\subsection{Solving ODE's and PDE’s.}
Solving ODE's and PDE's is often based on solution of leanear systems with sparse matrices over numbers or over polynomials.
One of the important class of sparse matrix is called quasiseparable. 
Any submatrix of quasiseparable matrix  entirely below or above the main diagonal has small rank.
 These quasiseparable matrices arise naturally in solving PDE’s for particle interaction with the Fast Multi-pole Method (FMM). 
 The efficiency of application of the block-recursive algorithm of the Bruhat decomposition to the quasiseparable 
 matrices is studied in \cite{Pernet-Stor-2017}. 
\section{Development of the recursive matrix agorithms in integral domain}

We can trace how developed the matrix recursive agorithms in integral domain,
which eventually led to the creation of modern algorithms.
There are several separate periods.
\subsection
{Algorithms for solution of a system of linear equations of size $n$ in an integral domain, which served as the basis for recursive algorithms} 
\ \\
{\it (1983)}  Forward and backward algorithm ($\sim n^3$) \cite{1983}.  \\
{\it (1989)}  One pass algorithm ($\sim \frac 23 n^3$) \cite{1989}.  \\ 
{\it  (1995)} Combined algoritm with upper left block of size $r$  ($\sim \frac 7{12}n^3$ for $r=\frac n 2$) \cite{1995}.\\
Really, this was already the first step of a recursive algorithm.
It was first discovered that when the matrix is divided into equal four blocks ($r=\frac n 2$), the least computational complexity is achieved.
Consequently, further dichotomous division of blocks can give the best algorithm.
It remains to prove several determinant identities that would allow us to do recursive calculations.

\subsection
{  Recursive algorithms for solution of a system of linear equations and for 
adjoint matrix computation in an integral domain without permutations 
}
\ \\
 {\it  (1997)}  Recursive algorithm for solution of a system of linear equations  \cite{1997}.\\
{\it (2000)}  Adjoint matrix computation (with 6 levels) \cite{2000}. \\
{\it  (2006)}  Adjoint matrix computation alternative algorithm (with 5 levels) \cite{2006}. \\
Now it remained to solve the problem of permutation of blocks and ensure the fulfillment of determinant identities.
\subsection
{ Main recursive algorithms for matrices in a domain}
\ \\
{\it  (2008)} Computation of adjoint and inverse matrices and the operator kernel in a domain \cite{2008}. \\
{\it (2010)} Bruhat and LEU decompositions in a feild  \cite{2010}. \\
{\it (2012)} Bruhat and LDU decompositions in a domain \cite{2012}, \cite{2013}. \\
Recursive algorithms for sparse matrices in commutative domains with the complexity of matrix multiplication are obtained.
The complexity of computing the matrix product for matrices of size $n$ we denote by $\sim n^\beta $.
\subsection{New achivements and new applications}
\ \\
{\it (2013)} It is proved that the LEU algorithm has the complexity $O(n^2r^{\beta-2})$ for rank $r$ matrices
\cite{Pernet-2013}.\\
{\it (2015)} New algorithms for Bruhat and LDU decompositions in a domain (alternative algorithm) \cite{2015}. \\
{\it (2017)} It is proved that the LEU algorithm has the complexity $O(n^2s^{\beta-2})$ for quasiseparable matrix. A matrix is called  
quasiseparable if any it's submatrix which entirely disposed below or above the main diagonal has small rank $s$, $s<<n$ \cite{Pernet-Stor-2017}.

\section{Recursive standard and strassen's matrix multiplication}

The graph of recursive algorithm for standard matrix multiplication is shown at Figure 1.

$$\begin{pmatrix} A_0 & A_1 \\ A_2 & A_3 \end{pmatrix} \times \begin{pmatrix} B_0 & B_1 
\\ B_2 & B_3 \end{pmatrix}
+\begin{pmatrix} C_0 & C_1 \\ C_2 & C_3 \end{pmatrix}=
\begin{pmatrix} D_0 & D_1 \\ D_2 & D_3 \end{pmatrix}$$ 
 
\begin{figure}
\caption
{Recursive standard matrix multiplication.} 
\bigskip

\centering\includegraphics[height=10cm, width=8.8cm]{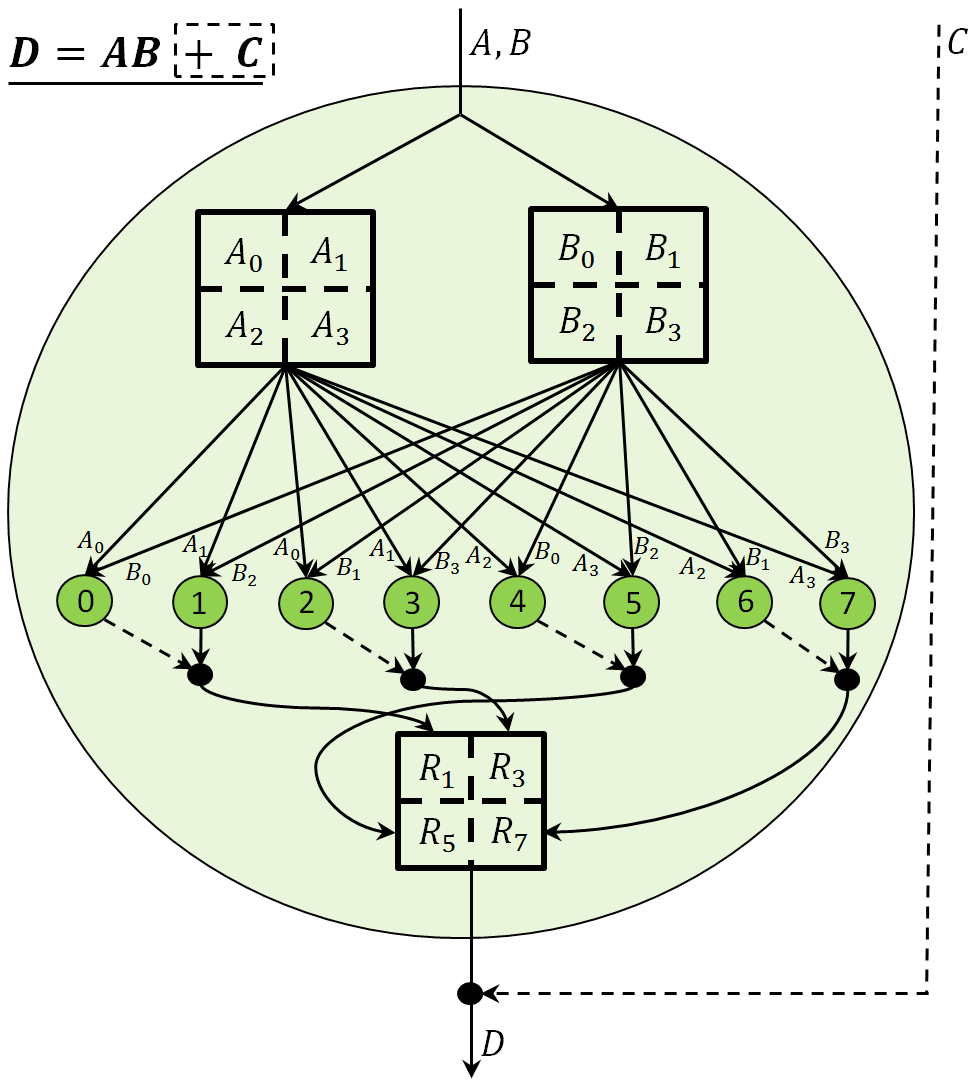}
\end{figure}
 
Number of operations for the standard algorithm is $\sim n^3$.

  The graph of the Strassen multiplication algorithm can be easily represented similarly.
The number of operations for this algorithm is $\sim n^{\log_2 7}$.

The algorithm for multiplying matrices on leaf tops should take into account the sparse matrix structure and compact storage form.

We note that there exists a boundary with respect to the density of the matrix, which separates the region of applicability of the Strassen multiplication. 
We note that there exists a theoretical boundary for   the
density of sparse matrix, which separates the region of efficient application of the Strassen algorithm of multiplication.
If the density of the matrix is below this boundary, then only standard multiplication is effective. This is due to the fact that the addition of 
blocks, which is performed in the Strassen algorithm, leads to an increase in the density of the matrix blocks  (see details in \cite{MVL}).

\section{ Recursive  Strassen's  matrix inversion }

If $\mathcal A=\begin{pmatrix} A_0 & A_1 \\ A_2 & A_3 \end{pmatrix}$,
 $\det({\mathcal A})\neq 0$  and $\det(A_0)\neq 0$ then
 $${  \mathcal A}^{-1}= 
\begin{pmatrix}\textbf{\emph{I}} & -A_0^{-1}A_1 \\ 0 & \textbf{\emph{I}} \end{pmatrix}  
\begin{pmatrix} \textbf{\emph{I}} & 0 \\ 0 & (A_3-A_2A_0^{-1}A_1)^{-1} \end{pmatrix} \times
$$
$$
\begin{pmatrix} \textbf{\emph{I}} & 0 \\ -A_2 & \textbf{\emph{I}} \end{pmatrix} 
\begin{pmatrix} A_0^{-1} & 0 \\ 0 & \textbf{\emph{I}} \end{pmatrix}
=\begin{pmatrix}
M_1  M_5-M_0 \ \  & M_1  M_4 \\ M_5 \ \  & M_4
\end{pmatrix}.
$$
We have denoted here  $M_0=-A_0^{-1}$,\  $M_1=M_0 A_1$,\  $M_2=A_2 M_0$,\  
$M_3=M_2 A_1$,\  $M_4=(A_3+M_3)^{-1}$,\  $M_5=-M_4  M_2$.

%

The graph of recursive Strassen's matrix inversion is shown on Figure 2.

\begin{figure}
\caption
{The graph of recursive Strassen's matrix inversion}
\bigskip

\centering\includegraphics[height=11cm, width=8.8cm]{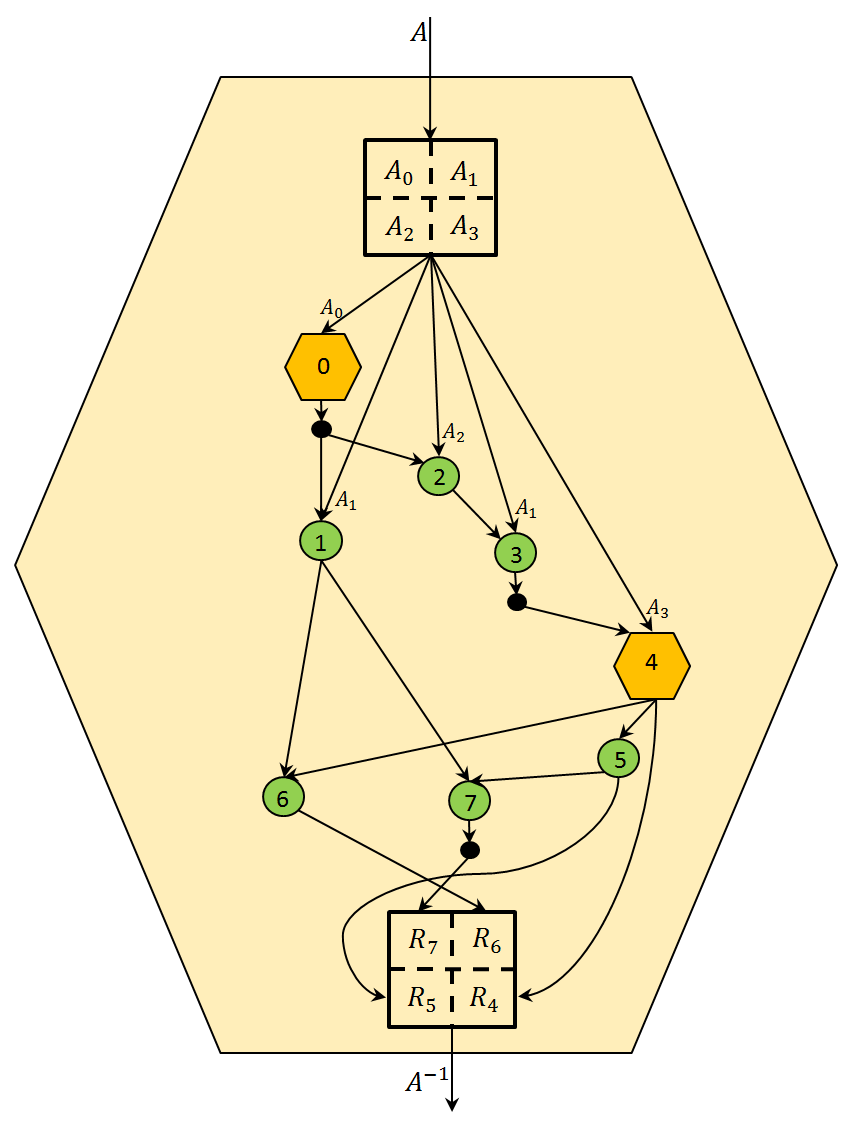}
\end{figure}

\section{ Recursive  inversion of triangular  matrix}
If $\mathcal A=\begin{pmatrix} A & 0 \\ B & C \end{pmatrix}$ is triangular matrix of order $2^k$ and
 $\det({\mathcal A})\neq 0$ then  
 $${  \mathcal A}^{-1}= 
\begin{pmatrix}  A ^{-1} & 0  \\-C^{-1} B A^{-1} & C^{-1}  \end{pmatrix}.
$$
 
\section{ Recursive Cholesky decomposition}
 
Let $\mathcal A=\begin{pmatrix} A_1 & A_2 \\ A_2^T & A_3 \end{pmatrix}$ be a positive definite symmetric matrix and 
$H=\begin{pmatrix} B & 0 \\ C & D \end{pmatrix}$ be a low triangle matrix with the property ${\mathcal A}= H H^T$.
The mapping 
$$ Chol: \ R^{n \times n}  \rightarrow (R^{n \times n}, R^{n \times n}),  $$
$$
  Chol({\mathcal A} )=(H, H^{-1}) 
$$
 is called an {\it Cholesky decomposition}.
 Let $ n = 2^k $, then you can used following recursive algorithm.

 \subsubsection{} 
 Let $Chol({  A_1} )=(B, B^{-1})$. 
 
 We can compute C=$A_2^T (B^{-1})^T$ and $F=A_3- C C^T$
  \subsubsection{} 
  Let $Chol({F} )=(D, D^{-1})$. 
  
 Then $H=\begin{pmatrix} B & 0 \\ C & D \end{pmatrix}$ and $H^{-1}=\begin{pmatrix} B^{-1} & 0 \\ -D^{-1} C B^{-1} & D^{-1} \end{pmatrix}$.
\section{Recursive computation of the adjoint matrix, kernel and determinant}
 
 We consider matrices over a commutative domain.
 
Semigroup $ P_n $ is formed by $n \times n$ matrices, which have the number of unit elements coincides with its rank, and
the remaining elements are zero. The semigroup $ D_n $ is formed by the diagonal
matrices:  $D_n \subset P_n$, $ | D_n | $ = $ 2 ^ n $. The identity matrix $ {\mathbf I} $ is a unit in $ D_n $ and in $ P_n $.
\
For each matrix $ E \in P_n $ we define diagonal matrices
$
I_E = E E ^ T \in D_n, 
J_E = E ^ T E \in D_n.
$ 
Also, we used the involution fanction on $ D_n: {\bar I} = {\mathbf I} - I$, $ \forall I\in D_n$ with the property ${\bar {\bar I}} =   I$.

 For the matrix $ E $, the matrix $ {\bar I_E} $ is a left annihilator, and the matrix $ {\bar J_E} $ is a right annihilator.
 So we can denote the set of echelon matrix of order $ n $:
$
S_n = \{ S\ | \exists E \in P_n, \ \exists d \in R \setminus 0: \ S = I_E S, \ dE = SJ_E \}.
$
In other words, $dE$ ($ E \in P_n,\ d\in R \setminus 0  $) is a block of echelon matrix $ S \in S_n $ with ${\rm rank}\ E={\rm rank}\ S$,    such that
the sets of zero rows of the matrices $ S $ and $ E $ coincide, and each nonzero column of the
matrix $ dE $ coincides with the same column of the matrix $ S $.
We write: $ E = E_S, S \in S_n $. 

Below we will use
  such notation for any matrix $S_{ij} \in S_n$ and $E_{ij} = (E_{ij})_{S_{ij}}$:
$$I_{ij}=E_{ij}E^{T}_{ij}, \bar{I}_{ij}=\textbf{\emph{I}}-I_{ij},
Y_{ij}=E^{T}_{ij}S_{ij}-d_{ij}\textbf{\emph{I}},\ \ i,j\in {1,2}.  $$

The mapping 
$$ A_{ext}: \ R^{n \times n} \times (R \backslash 0) \rightarrow (R^{n \times n})^3 \times (R \backslash 0), $$
$$
 A_ {ext} (M, d_0)=(A, S, E_S, d) 
$$
for $ n = 2^k $ is called an {\it extended adjoint mapping} of the pair ($ M, d_0 $) if it is defined recursively as follows.

For  $M=0$ we define
$
A_{ext}(0,d_0)=(d_0 I,0,0,d_0).
$
 
For $k=0$ and $M=a\neq 0$ we define
$
A_{ext}(a,d_0)=(d_0,a,a,a).
$

In all other cases, we split the matrix $ M $ into four equal blocks 
$M=\begin{pmatrix} M_{11} & M_{12} \\ M_{21} & M_{22} \end{pmatrix}$
\subsubsection{} 
Let
$A_{ext}(M_{11},d_{0})=(A_{11},S_{11},E_{11},d_{11}).$
 
We denote
$M^{1}_{12}= {A_{11}M_{12}}/{d_{0}}$,
$M^{1}_{21}=- {M_{21}Y_{11}}/{d_{0}}$,
$M^{1}_{22}= {M_{22}d_{11}-M_{21}E^{T}_{11}M^{1}_{12}}/{d_{0}}$.
\subsubsection{}
Let
$A_{ext}(\bar{I}_{11}M^{1}_{12},d_{11})=(A_{12},S_{12},E_{12},d_{12})$.
\subsubsection{}
Let
$A_{ext}(M^{1}_{21},d_{11})=(A_{21},S_{21},E_{21},d_{21}).$
 
We denote
$M^{2}_{22}=- {A_{21}M^{1}_{22}Y_{12}}/{(d_{11})^{2}}$, 
$d_{s}= {d_{21}d_{12}}/{d_{11}}.$
\subsubsection{}
Let $A_{ext}(\bar{I}_{21}M^{2}_{22},d_{s})=(A_{22},S_{22},E_{22},d_{22}).$
 
We denote
$M^{2}_{11}=-{S_{11}Y_{21}}/{d_{11}}$, $M^{2}_{12}=$
$$ ({ ( {\frac{S_{11}E^{T}_{21}A_{21}}{d_{11}}M^{1}_{22}-I_{11}M^{1}_{12}d_{21}} )/{d_{11}} *Y_{12}+S_{12}d_{21}})/{d_{11}},$$
$M^{3}_{12}=- {M^{2}_{12}Y_{22}}/{d_{s}}$, 
$M^{3}_{22}=S_{22}- {I_{21}M^{2}_{22}Y_{22}}/{d_{s}}$, 
 $$A^{1}=A_{12}A_{11}, \
L= ( {A^{1}-({I_{11}M^{1}_{12}E^{T}_{12}A^{1}})/{d_{11}}})/{d_{11}}*d_{22},$$
$$A^{2}=A_{22}A_{21}, \
P=({A^{2}-({I_{21}M^{2}_{22}E^{T}_{22}A^{2}})/{d_{s}}})/{d_{21}},$$
$F=-({( {S_{11}E^{T}_{21}A_{21}})/{d_{11}}*d_{22}+({M^{2}_{12}E^{T}_{22}A^{2}})/{d_{s}}})/{d_{21}}$,
\ \ \ 
$G=-({ ({M_{21}E^{T}_{11}A_{11}})/{d_{0}}*d_{12}+({M^{1}_{22}E^{T}_{12}A^{1}})/{d_{11}}})/{d_{11}},$
$$
A=\begin{pmatrix} (L+FG) /{d_{12}} & F \\ (PG)/d_{12} & P \end{pmatrix},
\
S=\begin{pmatrix} ({M^{2}_{11}d_{22}})/{d_{21}} & M^{3}_{12} \\[1.0ex]  
({S_{21}d_{22}})/{d_{21}} & M^{3}_{22} \end{pmatrix},
$$
$E=\begin{pmatrix} E_{11} & E_{12} \\ E_{21} & E_{22} \end{pmatrix},\ 
d=d_{22}.
$
$$ \hbox{Then   }
A_{ext}(M,d_{0})=(A,S,E,d).
$$
{{\it Sentence.} {The   map 
$A_{ext} (M, 1)= (A, S, E, d)$
defines an extended adjoint matrix 
$ A $, an echelon matrix $ S $, and a matrix $ E_S $ such that $ AM = S $ and $ dE = S J_E $ }
 \cite{2008}.}
The graph of extended adjoint map is shown at Figure 3.
 
\begin{figure}
\caption
{The graph of recursive computation of adjoint matrix and kernel.}
\bigskip

\centering\includegraphics[height=15cm, width=8.8cm]{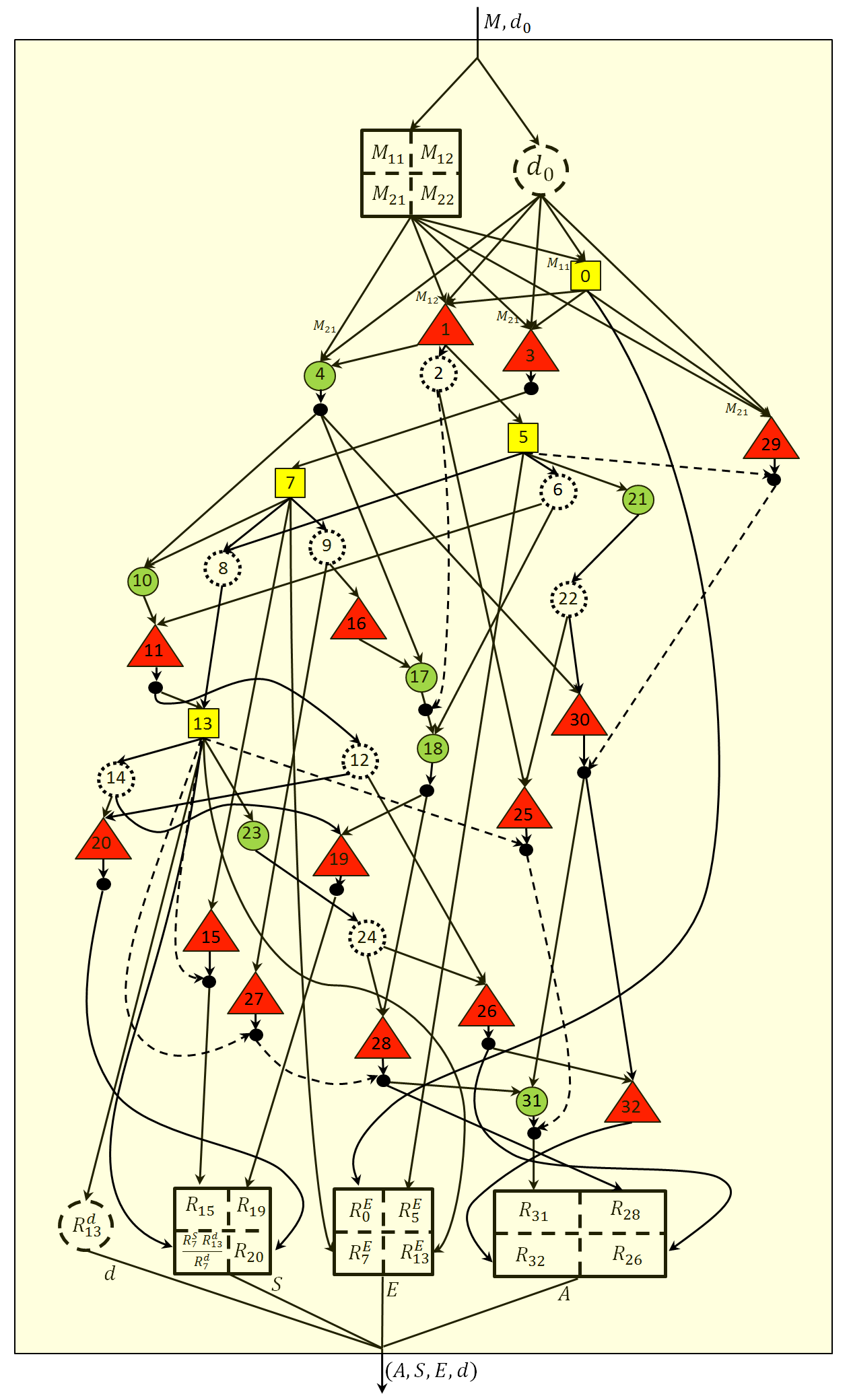}
\end{figure}

\section{Results of experiments with matrix recursive algorithms on a cluster with distributed memory}

The block-recursive matrix algorithms   require 
a special approachs to managing parallel programs.  
One approach to the cluster computations management is a scheme with one dispatcher.

We consider another scheme of cluster menagement. It is a scheme with multidispatching, when each involved computing
module has its own dispatch thread and several processing threads.
Each processor, along with its subtask, receives a list of slave processors. During the computation, 
this list changes when new processors are added or when some of these processors complete their subtasks 
 \cite{ilchenko13}, \cite{ilchenko15}. 

\subsection{Scalability}

We have done the experiments using the supercomputer MVS-10P based on RSC Tornado architecture.
 It is a 10 Petaflops supercomputing system at the Joint Supercomputer Center (JSCC) of the Russian Academy of Science (RAS).

We demonstrate the results of experiments with parallel programms on the base of a scheme with multidispatching, using C++ and OpenMPI.
 We demonstrate a scalability of these programs. To do this, we plot the graph of efficiency factor. 

For an ``ideal parallel program'' the product of the time $t_n$ for solving the computational problem by the number  $n$ 
of cores  in the computational cluster must remain constant: 
$$ t_n n = const.$$ 
So the value 
$$f=\frac{  t_n n}{ t_k k}  100\% $$
can be taken as the ``efficiency factor'' of the n-cores with respect to the k-cores.
 
In order to investigate how the efficiency factor changes with increasing number of cores for a given program, 
it is possible to conduct a series of experiments with different number of cores in the cluster.
If the program has this coefficient above 50\% for some range of number of cores, 
then we conside that it has good scalability in this range.
We conducted several series of experiments and obtained graphs 
that show how the efficiency factor varies. In all experiments, except Strassen's matrix inversion, 
we took matrices with 15 bit integers.

%
 
For the algorithm of recursive Strassen's matrix inversion, we took a dense matrix with double-precision numbers.
Figures 4 and 5 show the results of experiments with dence matrices of sizes 8000x8000 and 16000x16000, correspondingly.
For a cluster having 200 cores, the efficiency coefficient is equal to 51\% for a matrix size of 8000x8000 
and 73\% for a matrix size of 16000x16000. 

In Figure 6, the efficiency is shown for a recursive algorithm for computing the adjoint matrix and the kernel,
when the number of cores in a cluster changes from 8 to 400. We took arbitrary dense matrices with size 8000x8000.
For a cluster having 200 cores, the efficiency coefficient is equal to 66\%.  
Efficiency coefficient drops to 44\% when the number of cores reaches 400.

Figures 4-- 7 demonstrate that the larger the size of the matrix, the better
the efficiency coefficient remains with increasing cluster sizes.

\subsection{Comparison of the calculation time}
  The next three figures show how the calculation time varies with
  the growth of the matrix sizes for different algorithms and for different sparseness of the matrices.

We investigate the algorithm of computing  the adjoint matrix and the kernel (see the algorithm in Figure 3).
Experiments with two different versions of this algorithm  are compared in Figures 7 and 8. 
 These experiments were carried out for integer matrices on a cluster
 with 100 cores. We compared a standard algorithm and an algorithm 
 in which the Chinese remainder theorem (CRT) was applied. 
 It is well known that the application of the Chinese remainder theorem makes it possible to reduce 
 the number of operations in such algorithms by $\sim n$ times, where $n$ is the size of the matrix, if 
 the standard algorithm for multiplying integers is used. This is true for dense matrices.
 
 Figure 7 shows the results of experiments for dense matrices. For a matrix of size 2500x2500, the CRT
 algorithm is about twice as fast as a standard algorithm.

Figure 8 shows the results of experiments for sparse matrices that have a density of 1 percent. 
For a matrix of size 2500x2500, the CRT algorithm is approximately twice as slow as the standard algorithm.

We see that for very sparse matrices, the CRT algorithm should not be used. 

\section{Results of experiments with sequential programs: comparison with Mathematica and MAPLE}
 An experimental comparison of the sequential recursive algorithm for computing the adjoint matrix 
 with similar programs in Mathematica and MAPLE systems is shown in Figure 9.
  
 We compared our algorithms with Mathematica 11 and MAPLE 2015.
For comparison, we took random dense integer matrices and performed calculations with identical matrices in 4 programs.
The best time of calculations was demonstrated by sequential recursive algorithm. For example, for 600x600 matrices, 
it is twice as fast as Mathematica 11 and 7 times faster than MAPLE 2015. 
A slightly worse calculation time was shown by
the fourth program. This is the parallel program based on recursive algorithm that was run on a single processor.
 
\section{Conclusion}
 
 The algorithms we discussed were used in the cloud computing system of the computer algebra Math Partner.
You can used this cloud system on the websites http://mathpar.cloud.unihub.ru.
 
 Functions that correspond to these algorithms can be called by the operators {\it adjoint}, {\it kernel}, {\it det}, {\it LDU}, {\it BruhatDecomposition}.
  
  The algorithms that have been discussed above have a wide application. Therefore, it would be important to have such a public site where users could perform 
  parallel computing to solve their specific tasks using a large number of processors.
  
 
\section{Acknowledgment}

The authors are grateful to the Ivannikov Institute for System Programming of the Russian Academy of Science for hosting the cloud computer algebra system Math Partner and to
the Joint Supercomputer Center  of the Russian Academy of Science
   for providing the ability to perform calculations on the supercomputer MVS-10P. 
  
 
%
%
%
%
%

\begin{figure}
\caption
{Recursive Strassen matrix inversion, size=8000x8000.}
\bigskip

\centering\includegraphics[height=8cm, width=8cm]{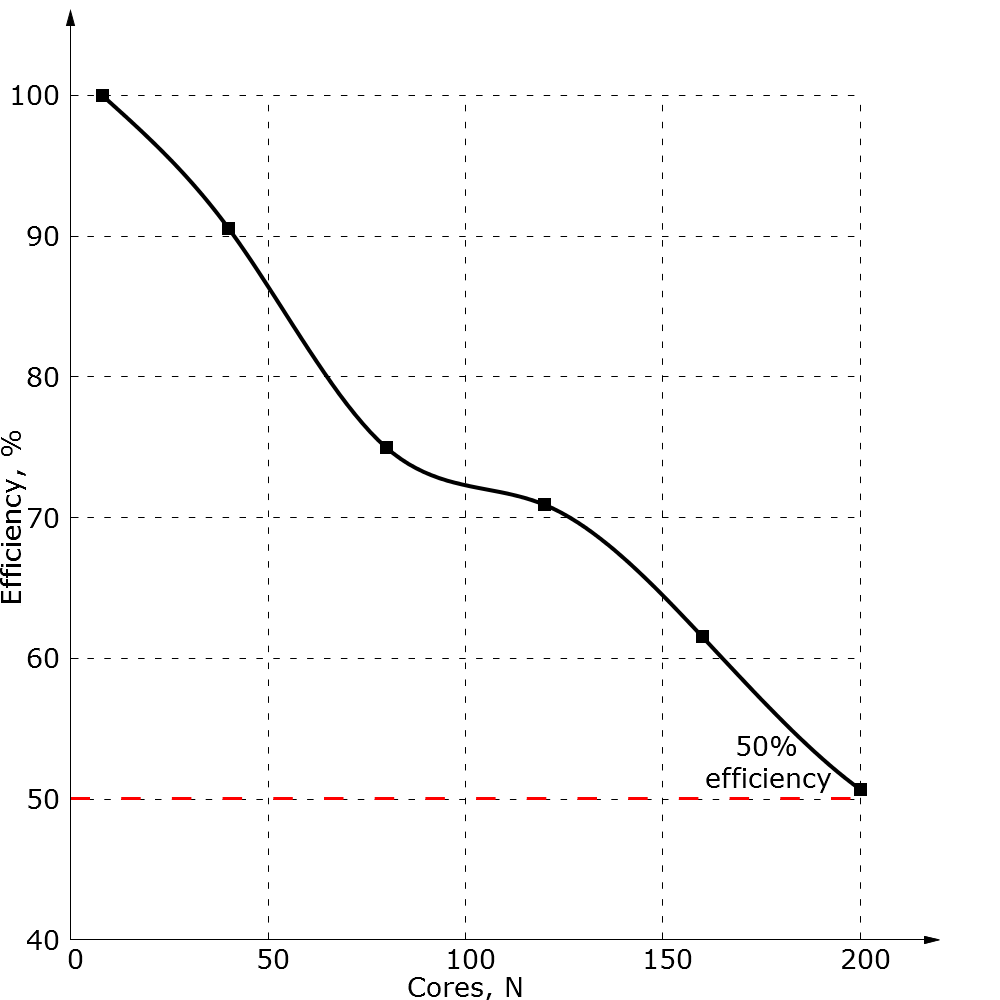}
\end{figure} 

\begin{figure}
\caption
{Recursive Strassen matrix inversion, size=16000x16000.}
\bigskip

\centering\includegraphics[height=10cm, width=8.8cm]{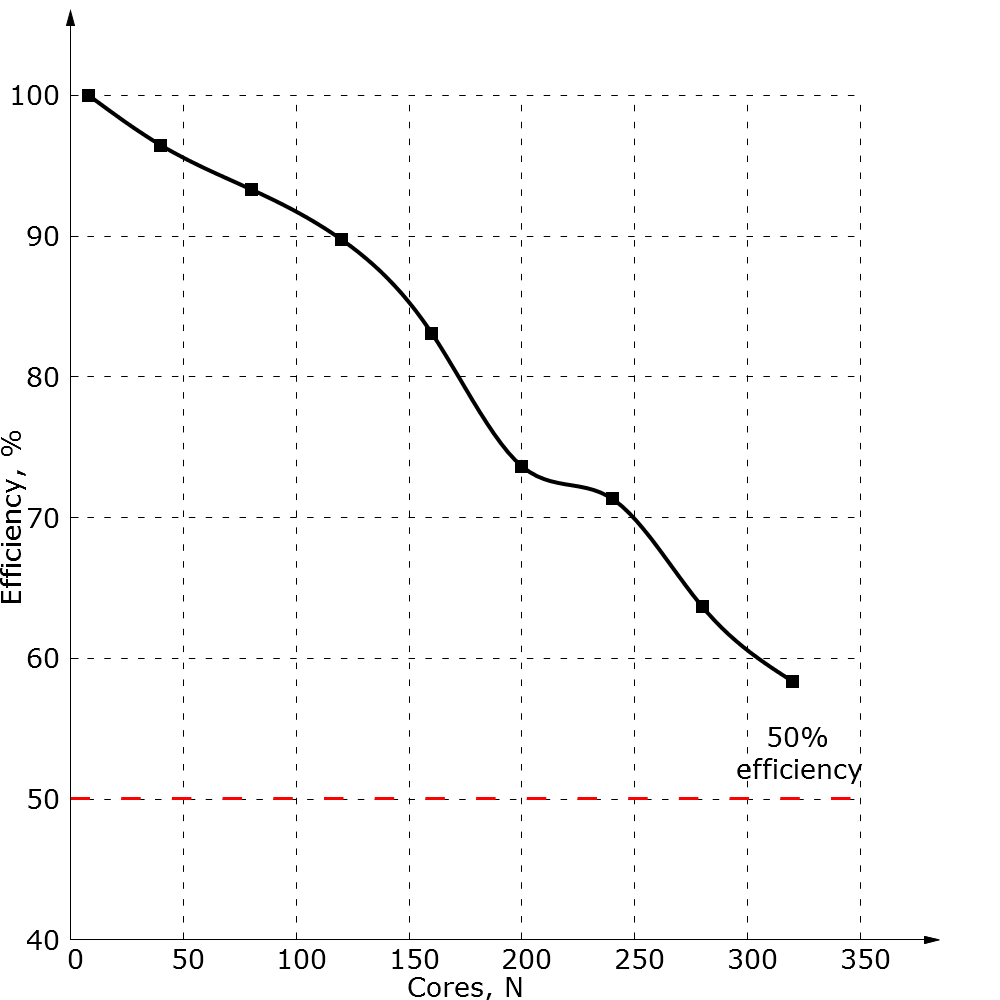}
\end{figure}
 
\begin{figure}
\caption
{Recursive algorithm for compution of adjoint matrix and kernel, size=8000x8000.}
\bigskip
 
\centering\includegraphics[height=8cm, width=8cm]{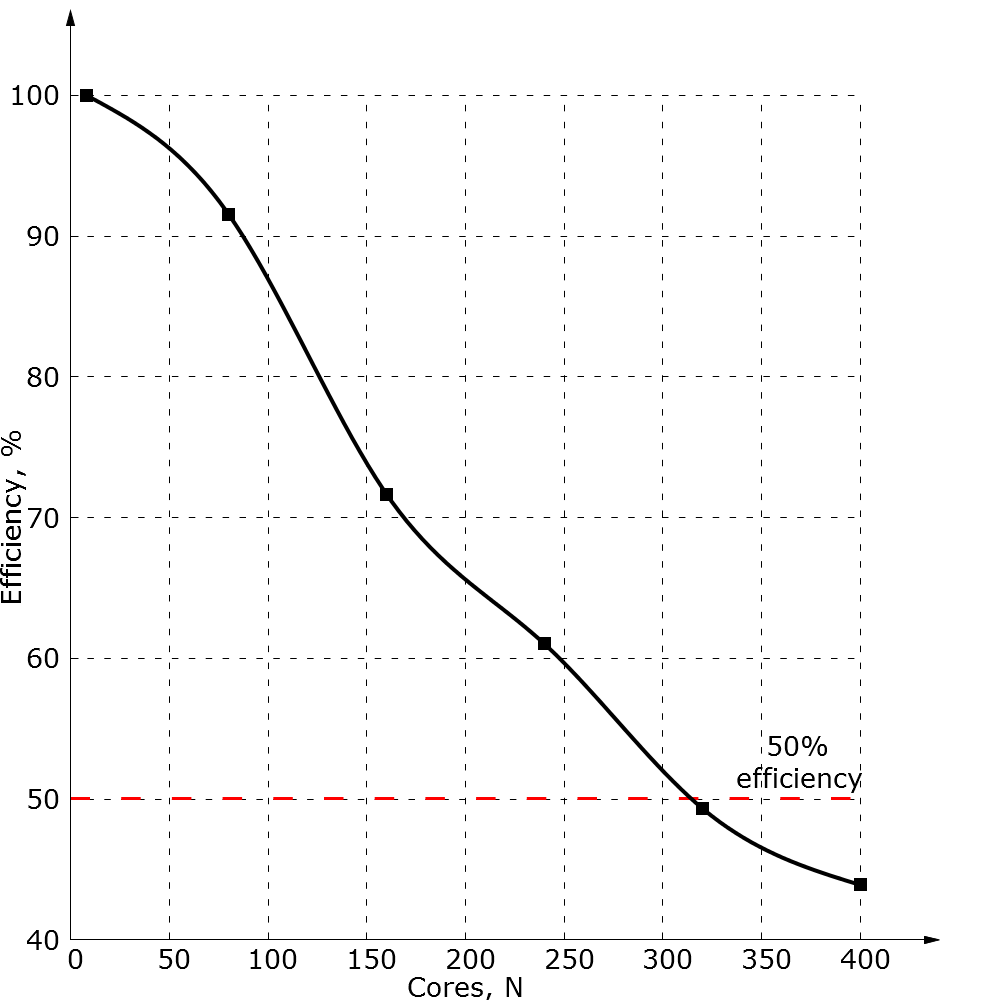}
\end{figure}
  
\begin{figure} 
\caption
{Recursive algorithm for compution of adjoint matrix and kernel on the cluster with 100 cores,
comparison of a standard algorithm and CRT algorithm for dense matrices.}
\bigskip

\centering\includegraphics[height=10cm, width=8.8cm]{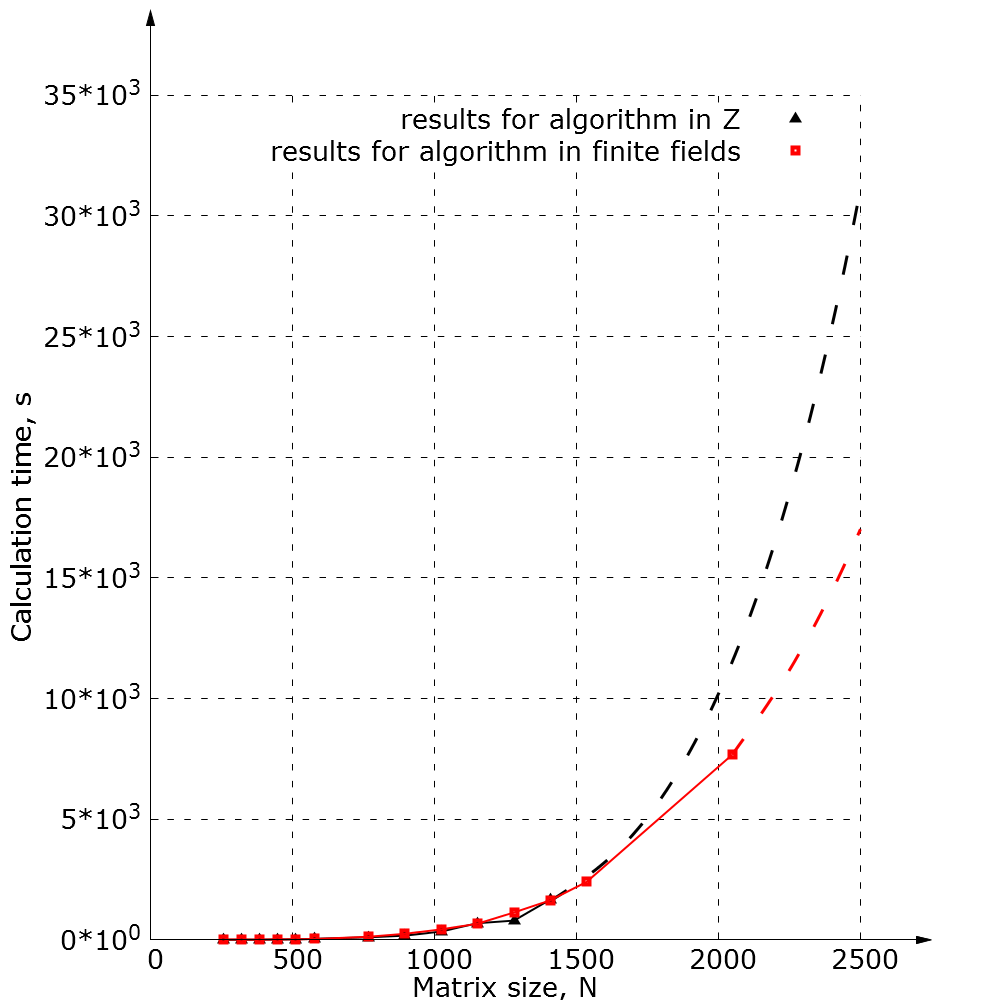}
\end{figure}

\begin{figure} 
\caption
{Comparison of a standard and CRT algorithms of compution of adjoint matrix and kernel 
for sparse matrices with 1\% of dencity.}
\bigskip

\centering\includegraphics[height=8cm, width=8cm]{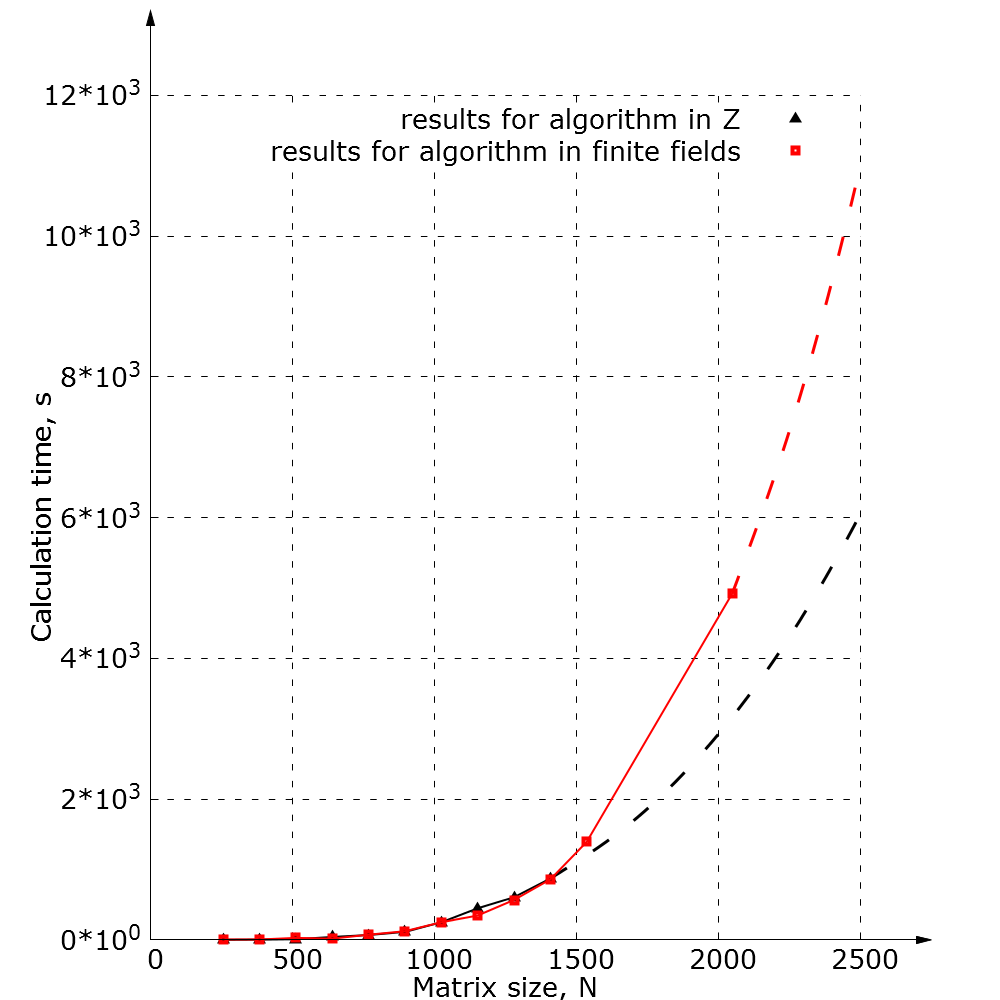}
\end{figure}

\begin{figure} 
 \caption{Comparison of the sequential recursive algorithm for computing the adjoint matrix and the 
 kernel with in Mathematica and MAPLE.}  
  \bigskip
  
\centering\includegraphics[height=10cm, width=8.8cm]{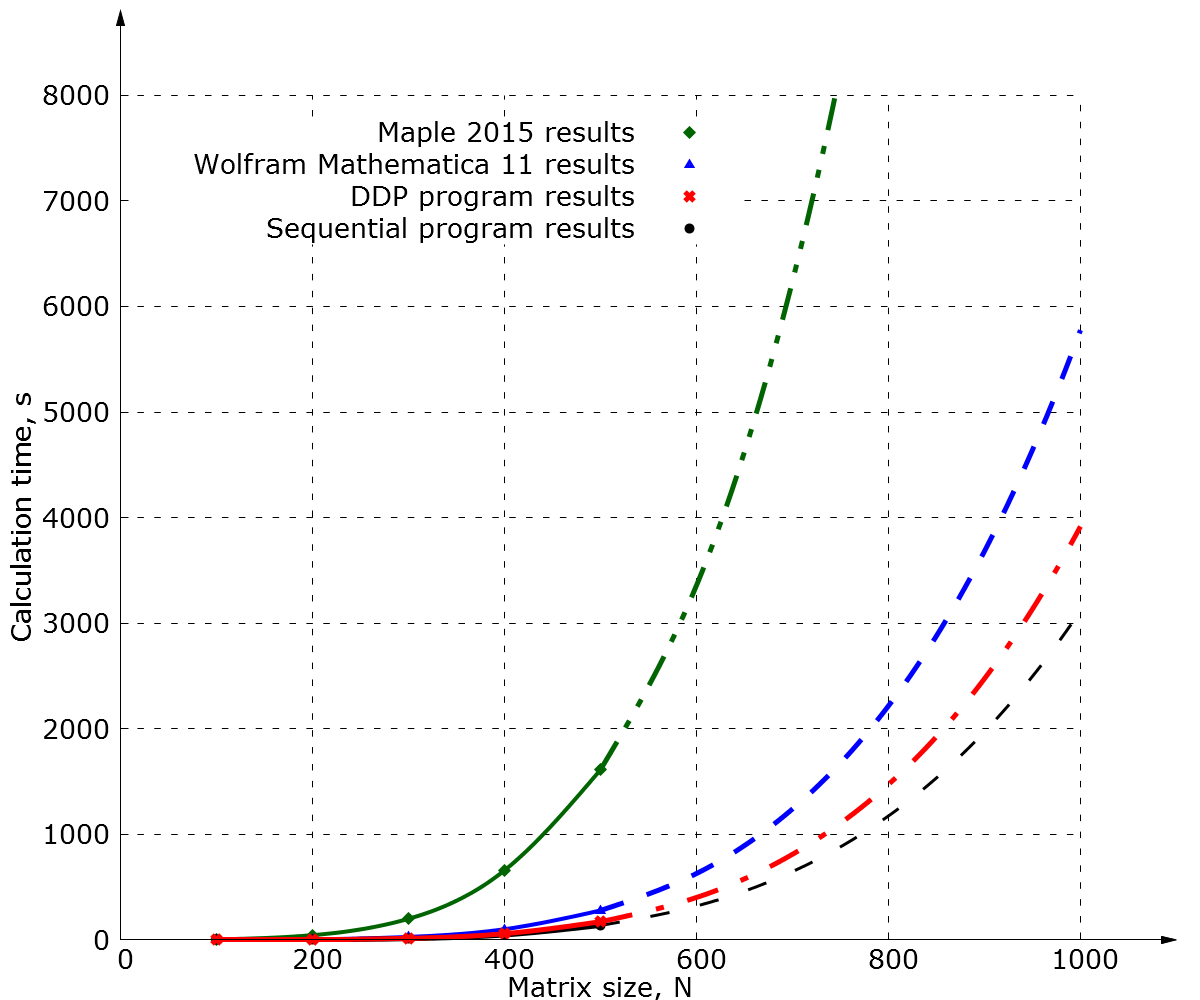}
 \end{figure}

\eject



\begin{thebibliography}{00}

\bibitem    
{Dongarra-2016}  J. Dongarra {``With Extrim Scale Computing the Rules Have Changed'',} in Mathematical Software. ICMS 2016, 5th International Congress, 
Proceedings (G.-M. Greuel, T. Koch, P. Paule, A. Sommese, eds.), Springer, LNCS, vol. 9725, pp. 3--8, 2016.

\bibitem{MVL}
G.I.Malashonok, Y.D.Valeev, A.O. Lapaev, 
``On the choice of a multiplication algorithm for polynomials and polynomial matrices'', 
J. Math Sci.  vol. 168, no. 3, pp. 398--416,  2010.

\bibitem 
{Strassen-1969}  { V. Strassen, `` Gaussian Elimination is not optimal,''} Numerische Mathematik. vol. 13, issue 4,
 pp. 354--356, 1969.

\bibitem 
{Bunch-Hopkroft-1974}
  J. Bunch,  J. Hopkroft, {``Triangular factorization and inversion by fast matrix
multiplication,''} Mat. Comp. vol.  28, pp. 231--236, 1974.

\bibitem{1983}
G.I.Malashonok, {``Solution of a system of linear equations in an integral domain, ''}  USSR J. of Comput. Math. and Math. Phys.,vol. 23, no. 6, pp. 1497--1500, 1983.

\bibitem{1989}
G.I.Malashonok, {``Algorithms for the solution of systems of linear equations in commutative rings. ''}
Effective methods in Algebraic Geometry, Progr. Math.,vol.  94, 
Birkhauser Boston, Boston, MA, 1991, pp. 289--298, 1991.
 
\bibitem{1995}
G.I. Malaschonok. {``Algorithms for computing determinants in commutative rings. ''}  
  Discrete Math. Appl., Vol. 5, no. 6, pp. 557--566, 1995.


\bibitem{1997}  G.I.Malashonok, {``Recursive Method for the Solution of Systems of Linear Equations,''} 
Computational Mathematics. A. Sydow Ed, Proceedings of the 15th IMACS World Congress, vol. I, (Berlin, August 1997), 
Wissenschaft \& Technik Verlag, Berlin, pp. 475--480, 1997.

\bibitem{2000} G. Malashonok, {`` Effective Matrix Methods in Commutative Domains''}, Formal Power Series and Algebraic Combinatorics, 
Springer, Berlin, pp. 506--517, 2000.

\bibitem{2002} G. Malashonok,  {`` Matrix computational methods in commutative rings,''} Tambov: Tambov State University,    2002.

\bibitem{2006}
 A.G.Akritas, G.I. Malashonok,  {`` Computation of Adjoint Matrix,''} Computational Science, ICCS 2006, LNCS 3992, Springer, Berlin, pp. 486--489, 2006.

\bibitem{2008} 
G. Malashonok,   {``On computation of kernel of operator acting in a module''}  
[Tambov University Reports. Series: Natural and Technical Sciences], vol.~13, issue~1, pp. 129--131, 2008.
 
\bibitem{2010}   
G. Malashonok,   {``Fast Generalized Bruhat Decomposition,''} Computer Algebra in Scientific Computing, LNCS 6244, Springer, Berlin,  pp. 194--202, 2010.

\bibitem{2012}
G. Malashonok,  {``On fast generalized Bruhat decomposition in the domains,''} Tambov University Reports. Series: Natural and Technical Sciences, vol. 17, issue 2, pp. 544--551, 2012.

\bibitem{2013} 
G. Malashonok,   {``Generalized Bruhat decomposition in commutative domains,''} Computer Algebra in Scientific Computing, CASC'2013, 
LNCS 8136, Springer, Heidelberg, 2013, pp. 231--242, 2013.

\bibitem{2015} 
G. Malashonok,   A. Scherbinin, {``Triangular Decomposition of Matrices in a Domain,''} Computer Algebra in Scientific Computing, 
LNCS 9301, Springer, Switzerland, 2015, pp. 290--304, 2015.


\bibitem{Paul-2001}
R.P. Clayton, {``Fundamentals of Electric Circuit Analysis,''}  
Mercer University \& University of Kentucky, John Wiley \& Sons, Inc., 2001.

\bibitem{Rosenbrock1967} 
Rosenbrock, H.H. {`` Transformation of linear constant system equations,''} Proc. I.E.E. vol. 114, pp. 541--544, 1967.

\bibitem{1999} 
Faugere, J.-C.  {``A new efficient algorithm for computing Gr\"obner bases (F4)''} . Journal of Pure and Applied Algebra. Elsevier Science, vol. 139, no. 1, pp. 61--88, 1999.

\bibitem{Pernet-2013}
J.-G. Dumas, C. Pernet, Z. Sultan,  {`` Simultaneous computation of the row and column
rank profiles,''} In: Kauers, M. (Ed.), Proc. ISSAC’13. ACM Press, pp. 181--188, 2013.

\bibitem{Pernet-Stor-2017} Pernet C., Storjohann A. {`` Time and space efficient generators for quasiseparable
matrices,''} Journal of Symbolic Computation, vol. 85, no. 2, pp. 224--246, 2018.
 
\bibitem{ilchenko13} Ilchenko E.A.  {``An algorithm for the decentralized control of parallel computing process,''} Tambov University Reports, series: Natural and Technical Sciences, 
vol. 18, no. 4, pp. 1198-1206, 2013.

\bibitem{ilchenko15} Ilchenko E.A. {``About effective methods parallelizing block recursive algorithms,''}  Tambov University Reports. series: Natural and Technical Sciences,  
vol. 20, no. 5, pp. 1173--1186, 2015.



\end{thebibliography}
\end{document}